\documentclass[superscriptaddress,prd,aps,preprint]{revtex4}

\usepackage{graphicx,amsfonts,amsmath}

\begin{document}
\immediate\write16{<<WARNING: LINEDRAW macros work with emTeX-dvivers
                    and other drivers supporting emTeX \special's
                    (dviscr, dvihplj, dvidot, dvips, dviwin, etc.) >>}

\newdimen\Lengthunit       \Lengthunit  = 1.5cm
\newcount\Nhalfperiods     \Nhalfperiods= 9
\newcount\magnitude        \magnitude = 1000

\catcode`\*=11
\newdimen\L*   \newdimen\d*   \newdimen\d**
\newdimen\dm*  \newdimen\dd*  \newdimen\dt*
\newdimen\a*   \newdimen\b*   \newdimen\c*
\newdimen\a**  \newdimen\b**
\newdimen\xL*  \newdimen\yL*
\newdimen\rx*  \newdimen\ry*
\newdimen\tmp* \newdimen\linwid*

\newcount\k*   \newcount\l*   \newcount\m*
\newcount\k**  \newcount\l**  \newcount\m**
\newcount\n*   \newcount\dn*  \newcount\r*
\newcount\N*   \newcount\*one \newcount\*two  \*one=1 \*two=2
\newcount\*ths \*ths=1000
\newcount\angle*  \newcount\q*  \newcount\q**
\newcount\angle** \angle**=0
\newcount\sc*     \sc*=0

\newtoks\cos*  \cos*={1}
\newtoks\sin*  \sin*={0}

\catcode`\[=13

\def\rotate(#1){\advance\angle**#1\angle*=\angle**
\q**=\angle*\ifnum\q**<0\q**=-\q**\fi
\ifnum\q**>360\q*=\angle*\divide\q*360\multiply\q*360\advance\angle*-\q*\fi
\ifnum\angle*<0\advance\angle*360\fi\q**=\angle*\divide\q**90\q**=\q**
\def\sgcos*{+}\def\sgsin*{+}\relax
\ifcase\q**\or
 \def\sgcos*{-}\def\sgsin*{+}\or
 \def\sgcos*{-}\def\sgsin*{-}\or
 \def\sgcos*{+}\def\sgsin*{-}\else\fi
\q*=\q**
\multiply\q*90\advance\angle*-\q*
\ifnum\angle*>45\sc*=1\angle*=-\angle*\advance\angle*90\else\sc*=0\fi
\def[##1,##2]{\ifnum\sc*=0\relax
\edef\cs*{\sgcos*.##1}\edef\sn*{\sgsin*.##2}\ifcase\q**\or
 \edef\cs*{\sgcos*.##2}\edef\sn*{\sgsin*.##1}\or
 \edef\cs*{\sgcos*.##1}\edef\sn*{\sgsin*.##2}\or
 \edef\cs*{\sgcos*.##2}\edef\sn*{\sgsin*.##1}\else\fi\else
\edef\cs*{\sgcos*.##2}\edef\sn*{\sgsin*.##1}\ifcase\q**\or
 \edef\cs*{\sgcos*.##1}\edef\sn*{\sgsin*.##2}\or
 \edef\cs*{\sgcos*.##2}\edef\sn*{\sgsin*.##1}\or
 \edef\cs*{\sgcos*.##1}\edef\sn*{\sgsin*.##2}\else\fi\fi
\cos*={\cs*}\sin*={\sn*}\global\edef\gcos*{\cs*}\global\edef\gsin*{\sn*}}\relax
\ifcase\angle*[9999,0]\or
[999,017]\or[999,034]\or[998,052]\or[997,069]\or[996,087]\or
[994,104]\or[992,121]\or[990,139]\or[987,156]\or[984,173]\or
[981,190]\or[978,207]\or[974,224]\or[970,241]\or[965,258]\or
[961,275]\or[956,292]\or[951,309]\or[945,325]\or[939,342]\or
[933,358]\or[927,374]\or[920,390]\or[913,406]\or[906,422]\or
[898,438]\or[891,453]\or[882,469]\or[874,484]\or[866,499]\or
[857,515]\or[848,529]\or[838,544]\or[829,559]\or[819,573]\or
[809,587]\or[798,601]\or[788,615]\or[777,629]\or[766,642]\or
[754,656]\or[743,669]\or[731,681]\or[719,694]\or[707,707]\or
\else[9999,0]\fi}

\catcode`\[=12

\def\GRAPH(hsize=#1)#2{\hbox to #1\Lengthunit{#2\hss}}

\def\Linewidth#1{\global\linwid*=#1\relax
\global\divide\linwid*10\global\multiply\linwid*\mag
\global\divide\linwid*100\special{em:linewidth \the\linwid*}}

\Linewidth{.4pt}
\def\sm*{\special{em:moveto}}
\def\sl*{\special{em:lineto}}
\let\moveto=\sm*
\let\lineto=\sl*
\newbox\spm*   \newbox\spl*
\setbox\spm*\hbox{\sm*}
\setbox\spl*\hbox{\sl*}

\def\mov#1(#2,#3)#4{\rlap{\L*=#1\Lengthunit
\xL*=#2\L* \yL*=#3\L*
\xL*=\xscale\xL* \yL*=\yscale\yL*
\rx* \the\cos*\xL* \tmp* \the\sin*\yL* \advance\rx*-\tmp*
\ry* \the\cos*\yL* \tmp* \the\sin*\xL* \advance\ry*\tmp*
\kern\rx*\raise\ry*\hbox{#4}}}

\def\rmov*(#1,#2)#3{\rlap{\xL*=#1\yL*=#2\relax
\rx* \the\cos*\xL* \tmp* \the\sin*\yL* \advance\rx*-\tmp*
\ry* \the\cos*\yL* \tmp* \the\sin*\xL* \advance\ry*\tmp*
\kern\rx*\raise\ry*\hbox{#3}}}

\def\lin#1(#2,#3){\rlap{\sm*\mov#1(#2,#3){\sl*}}}

\def\arr*(#1,#2,#3){\rmov*(#1\dd*,#1\dt*){\sm*
\rmov*(#2\dd*,#2\dt*){\rmov*(#3\dt*,-#3\dd*){\sl*}}\sm*
\rmov*(#2\dd*,#2\dt*){\rmov*(-#3\dt*,#3\dd*){\sl*}}}}

\def\arrow#1(#2,#3){\rlap{\lin#1(#2,#3)\mov#1(#2,#3){\relax
\d**=-.012\Lengthunit\dd*=#2\d**\dt*=#3\d**
\arr*(1,10,4)\arr*(3,8,4)\arr*(4.8,4.2,3)}}}

\def\arrlin#1(#2,#3){\rlap{\L*=#1\Lengthunit\L*=.5\L*
\lin#1(#2,#3)\rmov*(#2\L*,#3\L*){\arrow.1(#2,#3)}}}

\def\dasharrow#1(#2,#3){\rlap{{\Lengthunit=0.9\Lengthunit
\dashlin#1(#2,#3)\mov#1(#2,#3){\sm*}}\mov#1(#2,#3){\sl*
\d**=-.012\Lengthunit\dd*=#2\d**\dt*=#3\d**
\arr*(1,10,4)\arr*(3,8,4)\arr*(4.8,4.2,3)}}}

\def\clap#1{\hbox to 0pt{\hss #1\hss}}

\def\ind(#1,#2)#3{\rlap{\L*=.1\Lengthunit
\xL*=#1\L* \yL*=#2\L*
\rx* \the\cos*\xL* \tmp* \the\sin*\yL* \advance\rx*-\tmp*
\ry* \the\cos*\yL* \tmp* \the\sin*\xL* \advance\ry*\tmp*
\kern\rx*\raise\ry*\hbox{\lower2pt\clap{$#3$}}}}

\def\sh*(#1,#2)#3{\rlap{\dm*=\the\n*\d**
\xL*=\xscale\dm* \yL*=\yscale\dm* \xL*=#1\xL* \yL*=#2\yL*
\rx* \the\cos*\xL* \tmp* \the\sin*\yL* \advance\rx*-\tmp*
\ry* \the\cos*\yL* \tmp* \the\sin*\xL* \advance\ry*\tmp*
\kern\rx*\raise\ry*\hbox{#3}}}

\def\calcnum*#1(#2,#3){\a*=1000sp\b*=1000sp\a*=#2\a*\b*=#3\b*
\ifdim\a*<0pt\a*-\a*\fi\ifdim\b*<0pt\b*-\b*\fi
\ifdim\a*>\b*\c*=.96\a*\advance\c*.4\b*
\else\c*=.96\b*\advance\c*.4\a*\fi
\k*\a*\multiply\k*\k*\l*\b*\multiply\l*\l*
\m*\k*\advance\m*\l*\n*\c*\r*\n*\multiply\n*\n*
\dn*\m*\advance\dn*-\n*\divide\dn*2\divide\dn*\r*
\advance\r*\dn*
\c*=\the\Nhalfperiods5sp\c*=#1\c*\ifdim\c*<0pt\c*-\c*\fi
\multiply\c*\r*\N*\c*\divide\N*10000}

\def\dashlin#1(#2,#3){\rlap{\calcnum*#1(#2,#3)\relax
\d**=#1\Lengthunit\ifdim\d**<0pt\d**-\d**\fi
\divide\N*2\multiply\N*2\advance\N*\*one
\divide\d**\N*\sm*\n*\*one\sh*(#2,#3){\sl*}\loop
\advance\n*\*one\sh*(#2,#3){\sm*}\advance\n*\*one
\sh*(#2,#3){\sl*}\ifnum\n*<\N*\repeat}}

\def\dashdotlin#1(#2,#3){\rlap{\calcnum*#1(#2,#3)\relax
\d**=#1\Lengthunit\ifdim\d**<0pt\d**-\d**\fi
\divide\N*2\multiply\N*2\advance\N*1\multiply\N*2\relax
\divide\d**\N*\sm*\n*\*two\sh*(#2,#3){\sl*}\loop
\advance\n*\*one\sh*(#2,#3){\kern-1.48pt\lower.5pt\hbox{\rm.}}\relax
\advance\n*\*one\sh*(#2,#3){\sm*}\advance\n*\*two
\sh*(#2,#3){\sl*}\ifnum\n*<\N*\repeat}}

\def\shl*(#1,#2)#3{\kern#1#3\lower#2#3\hbox{\unhcopy\spl*}}

\def\trianglin#1(#2,#3){\rlap{\toks0={#2}\toks1={#3}\calcnum*#1(#2,#3)\relax
\dd*=.57\Lengthunit\dd*=#1\dd*\divide\dd*\N*
\divide\dd*\*ths \multiply\dd*\magnitude
\d**=#1\Lengthunit\ifdim\d**<0pt\d**-\d**\fi
\multiply\N*2\divide\d**\N*\sm*\n*\*one\loop
\shl**{\dd*}\dd*-\dd*\advance\n*2\relax
\ifnum\n*<\N*\repeat\n*\N*\shl**{0pt}}}

\def\wavelin#1(#2,#3){\rlap{\toks0={#2}\toks1={#3}\calcnum*#1(#2,#3)\relax
\dd*=.23\Lengthunit\dd*=#1\dd*\divide\dd*\N*
\divide\dd*\*ths \multiply\dd*\magnitude
\d**=#1\Lengthunit\ifdim\d**<0pt\d**-\d**\fi
\multiply\N*4\divide\d**\N*\sm*\n*\*one\loop
\shl**{\dd*}\dt*=1.3\dd*\advance\n*\*one
\shl**{\dt*}\advance\n*\*one
\shl**{\dd*}\advance\n*\*two
\dd*-\dd*\ifnum\n*<\N*\repeat\n*\N*\shl**{0pt}}}

\def\w*lin(#1,#2){\rlap{\toks0={#1}\toks1={#2}\d**=\Lengthunit\dd*=-.12\d**
\divide\dd*\*ths \multiply\dd*\magnitude
\N*8\divide\d**\N*\sm*\n*\*one\loop
\shl**{\dd*}\dt*=1.3\dd*\advance\n*\*one
\shl**{\dt*}\advance\n*\*one
\shl**{\dd*}\advance\n*\*one
\shl**{0pt}\dd*-\dd*\advance\n*1\ifnum\n*<\N*\repeat}}

\def\l*arc(#1,#2)[#3][#4]{\rlap{\toks0={#1}\toks1={#2}\d**=\Lengthunit
\dd*=#3.037\d**\dd*=#4\dd*\dt*=#3.049\d**\dt*=#4\dt*\ifdim\d**>10mm\relax
\d**=.25\d**\n*\*one\shl**{-\dd*}\n*\*two\shl**{-\dt*}\n*3\relax
\shl**{-\dd*}\n*4\relax\shl**{0pt}\else
\ifdim\d**>5mm\d**=.5\d**\n*\*one\shl**{-\dt*}\n*\*two
\shl**{0pt}\else\n*\*one\shl**{0pt}\fi\fi}}

\def\d*arc(#1,#2)[#3][#4]{\rlap{\toks0={#1}\toks1={#2}\d**=\Lengthunit
\dd*=#3.037\d**\dd*=#4\dd*\d**=.25\d**\sm*\n*\*one\shl**{-\dd*}\relax
\n*3\relax\sh*(#1,#2){\xL*=\xscale\dd*\yL*=\yscale\dd*
\kern#2\xL*\lower#1\yL*\hbox{\sm*}}\n*4\relax\shl**{0pt}}}

\def\shl**#1{\c*=\the\n*\d**\d*=#1\relax
\a*=\the\toks0\c*\b*=\the\toks1\d*\advance\a*-\b*
\b*=\the\toks1\c*\d*=\the\toks0\d*\advance\b*\d*
\a*=\xscale\a*\b*=\yscale\b*
\rx* \the\cos*\a* \tmp* \the\sin*\b* \advance\rx*-\tmp*
\ry* \the\cos*\b* \tmp* \the\sin*\a* \advance\ry*\tmp*
\raise\ry*\rlap{\kern\rx*\unhcopy\spl*}}

\def\wlin*#1(#2,#3)[#4]{\rlap{\toks0={#2}\toks1={#3}\relax
\c*=#1\l*\c*\c*=.01\Lengthunit\m*\c*\divide\l*\m*
\c*=\the\Nhalfperiods5sp\multiply\c*\l*\N*\c*\divide\N*\*ths
\divide\N*2\multiply\N*2\advance\N*\*one
\dd*=.002\Lengthunit\dd*=#4\dd*\multiply\dd*\l*\divide\dd*\N*
\divide\dd*\*ths \multiply\dd*\magnitude
\d**=#1\multiply\N*4\divide\d**\N*\sm*\n*\*one\loop
\shl**{\dd*}\dt*=1.3\dd*\advance\n*\*one
\shl**{\dt*}\advance\n*\*one
\shl**{\dd*}\advance\n*\*two
\dd*-\dd*\ifnum\n*<\N*\repeat\n*\N*\shl**{0pt}}}

\def\wavebox#1{\setbox0\hbox{#1}\relax
\a*=\wd0\advance\a*14pt\b*=\ht0\advance\b*\dp0\advance\b*14pt\relax
\hbox{\kern9pt\relax
\rmov*(0pt,\ht0){\rmov*(-7pt,7pt){\wlin*\a*(1,0)[+]\wlin*\b*(0,-1)[-]}}\relax
\rmov*(\wd0,-\dp0){\rmov*(7pt,-7pt){\wlin*\a*(-1,0)[+]\wlin*\b*(0,1)[-]}}\relax
\box0\kern9pt}}

\def\rectangle#1(#2,#3){\relax
\lin#1(#2,0)\lin#1(0,#3)\mov#1(0,#3){\lin#1(#2,0)}\mov#1(#2,0){\lin#1(0,#3)}}

\def\dashrectangle#1(#2,#3){\dashlin#1(#2,0)\dashlin#1(0,#3)\relax
\mov#1(0,#3){\dashlin#1(#2,0)}\mov#1(#2,0){\dashlin#1(0,#3)}}

\def\waverectangle#1(#2,#3){\L*=#1\Lengthunit\a*=#2\L*\b*=#3\L*
\ifdim\a*<0pt\a*-\a*\def\x*{-1}\else\def\x*{1}\fi
\ifdim\b*<0pt\b*-\b*\def\y*{-1}\else\def\y*{1}\fi
\wlin*\a*(\x*,0)[-]\wlin*\b*(0,\y*)[+]\relax
\mov#1(0,#3){\wlin*\a*(\x*,0)[+]}\mov#1(#2,0){\wlin*\b*(0,\y*)[-]}}

\def\calcparab*{\ifnum\n*>\m*\k*\N*\advance\k*-\n*\else\k*\n*\fi
\a*=\the\k* sp\a*=10\a*\b*\dm*\advance\b*-\a*\k*\b*
\a*=\the\*ths\b*\divide\a*\l*\multiply\a*\k*
\divide\a*\l*\k*\*ths\r*\a*\advance\k*-\r*\dt*=\the\k*\L*}

\def\arcto#1(#2,#3)[#4]{\rlap{\toks0={#2}\toks1={#3}\calcnum*#1(#2,#3)\relax
\dm*=135sp\dm*=#1\dm*\d**=#1\Lengthunit\ifdim\dm*<0pt\dm*-\dm*\fi
\multiply\dm*\r*\a*=.3\dm*\a*=#4\a*\ifdim\a*<0pt\a*-\a*\fi
\advance\dm*\a*\N*\dm*\divide\N*10000\relax
\divide\N*2\multiply\N*2\advance\N*\*one
\L*=-.25\d**\L*=#4\L*\divide\d**\N*\divide\L*\*ths
\m*\N*\divide\m*2\dm*=\the\m*5sp\l*\dm*\sm*\n*\*one\loop
\calcparab*\shl**{-\dt*}\advance\n*1\ifnum\n*<\N*\repeat}}

\def\arrarcto#1(#2,#3)[#4]{\L*=#1\Lengthunit\L*=.54\L*
\arcto#1(#2,#3)[#4]\rmov*(#2\L*,#3\L*){\d*=.457\L*\d*=#4\d*\d**-\d*
\rmov*(#3\d**,#2\d*){\arrow.02(#2,#3)}}}

\def\dasharcto#1(#2,#3)[#4]{\rlap{\toks0={#2}\toks1={#3}\relax
\calcnum*#1(#2,#3)\dm*=\the\N*5sp\a*=.3\dm*\a*=#4\a*\ifdim\a*<0pt\a*-\a*\fi
\advance\dm*\a*\N*\dm*
\divide\N*20\multiply\N*2\advance\N*1\d**=#1\Lengthunit
\L*=-.25\d**\L*=#4\L*\divide\d**\N*\divide\L*\*ths
\m*\N*\divide\m*2\dm*=\the\m*5sp\l*\dm*
\sm*\n*\*one\loop\calcparab*
\shl**{-\dt*}\advance\n*1\ifnum\n*>\N*\else\calcparab*
\sh*(#2,#3){\xL*=#3\dt* \yL*=#2\dt*
\rx* \the\cos*\xL* \tmp* \the\sin*\yL* \advance\rx*\tmp*
\ry* \the\cos*\yL* \tmp* \the\sin*\xL* \advance\ry*-\tmp*
\kern\rx*\lower\ry*\hbox{\sm*}}\fi
\advance\n*1\ifnum\n*<\N*\repeat}}

\def\*shl*#1{\c*=\the\n*\d**\advance\c*#1\a**\d*\dt*\advance\d*#1\b**
\a*=\the\toks0\c*\b*=\the\toks1\d*\advance\a*-\b*
\b*=\the\toks1\c*\d*=\the\toks0\d*\advance\b*\d*
\rx* \the\cos*\a* \tmp* \the\sin*\b* \advance\rx*-\tmp*
\ry* \the\cos*\b* \tmp* \the\sin*\a* \advance\ry*\tmp*
\raise\ry*\rlap{\kern\rx*\unhcopy\spl*}}

\def\calcnormal*#1{\b**=10000sp\a**\b**\k*\n*\advance\k*-\m*
\multiply\a**\k*\divide\a**\m*\a**=#1\a**\ifdim\a**<0pt\a**-\a**\fi
\ifdim\a**>\b**\d*=.96\a**\advance\d*.4\b**
\else\d*=.96\b**\advance\d*.4\a**\fi
\d*=.01\d*\r*\d*\divide\a**\r*\divide\b**\r*
\ifnum\k*<0\a**-\a**\fi\d*=#1\d*\ifdim\d*<0pt\b**-\b**\fi
\k*\a**\a**=\the\k*\dd*\k*\b**\b**=\the\k*\dd*}

\def\wavearcto#1(#2,#3)[#4]{\rlap{\toks0={#2}\toks1={#3}\relax
\calcnum*#1(#2,#3)\c*=\the\N*5sp\a*=.4\c*\a*=#4\a*\ifdim\a*<0pt\a*-\a*\fi
\advance\c*\a*\N*\c*\divide\N*20\multiply\N*2\advance\N*-1\multiply\N*4\relax
\d**=#1\Lengthunit\dd*=.012\d**
\divide\dd*\*ths \multiply\dd*\magnitude
\ifdim\d**<0pt\d**-\d**\fi\L*=.25\d**
\divide\d**\N*\divide\dd*\N*\L*=#4\L*\divide\L*\*ths
\m*\N*\divide\m*2\dm*=\the\m*0sp\l*\dm*
\sm*\n*\*one\loop\calcnormal*{#4}\calcparab*
\*shl*{1}\advance\n*\*one\calcparab*
\*shl*{1.3}\advance\n*\*one\calcparab*
\*shl*{1}\advance\n*2\dd*-\dd*\ifnum\n*<\N*\repeat\n*\N*\shl**{0pt}}}

\def\triangarcto#1(#2,#3)[#4]{\rlap{\toks0={#2}\toks1={#3}\relax
\calcnum*#1(#2,#3)\c*=\the\N*5sp\a*=.4\c*\a*=#4\a*\ifdim\a*<0pt\a*-\a*\fi
\advance\c*\a*\N*\c*\divide\N*20\multiply\N*2\advance\N*-1\multiply\N*2\relax
\d**=#1\Lengthunit\dd*=.012\d**
\divide\dd*\*ths \multiply\dd*\magnitude
\ifdim\d**<0pt\d**-\d**\fi\L*=.25\d**
\divide\d**\N*\divide\dd*\N*\L*=#4\L*\divide\L*\*ths
\m*\N*\divide\m*2\dm*=\the\m*0sp\l*\dm*
\sm*\n*\*one\loop\calcnormal*{#4}\calcparab*
\*shl*{1}\advance\n*2\dd*-\dd*\ifnum\n*<\N*\repeat\n*\N*\shl**{0pt}}}

\def\hr*#1{\L*=\xscale\Lengthunit\ifnum
\angle**=0\clap{\vrule width#1\L* height.1pt}\else
\L*=#1\L*\L*=.5\L*\rmov*(-\L*,0pt){\sm*}\rmov*(\L*,0pt){\sl*}\fi}

\def\shade#1[#2]{\rlap{\Lengthunit=#1\Lengthunit
\special{em:linewidth .001pt}\relax
\mov(0,#2.05){\hr*{.994}}\mov(0,#2.1){\hr*{.980}}\relax
\mov(0,#2.15){\hr*{.953}}\mov(0,#2.2){\hr*{.916}}\relax
\mov(0,#2.25){\hr*{.867}}\mov(0,#2.3){\hr*{.798}}\relax
\mov(0,#2.35){\hr*{.715}}\mov(0,#2.4){\hr*{.603}}\relax
\mov(0,#2.45){\hr*{.435}}\special{em:linewidth \the\linwid*}}}

\def\dshade#1[#2]{\rlap{\special{em:linewidth .001pt}\relax
\Lengthunit=#1\Lengthunit\if#2-\def\t*{+}\else\def\t*{-}\fi
\mov(0,\t*.025){\relax
\mov(0,#2.05){\hr*{.995}}\mov(0,#2.1){\hr*{.988}}\relax
\mov(0,#2.15){\hr*{.969}}\mov(0,#2.2){\hr*{.937}}\relax
\mov(0,#2.25){\hr*{.893}}\mov(0,#2.3){\hr*{.836}}\relax
\mov(0,#2.35){\hr*{.760}}\mov(0,#2.4){\hr*{.662}}\relax
\mov(0,#2.45){\hr*{.531}}\mov(0,#2.5){\hr*{.320}}\relax
\special{em:linewidth \the\linwid*}}}}

\def\vdot{\rlap{\kern-1.9pt\lower1.8pt\hbox{$\scriptstyle\bullet$}}}
\def\vtimes{\rlap{\kern-3pt\lower1.8pt\hbox{$\scriptstyle\times$}}}
\def\vDot{\rlap{\kern-2.3pt\lower2.7pt\hbox{$\bullet$}}}
\def\vTimes{\rlap{\kern-3.6pt\lower2.4pt\hbox{$\times$}}}

\def\arc(#1)[#2,#3]{{\k*=#2\l*=#3\m*=\l*
\advance\m*-6\ifnum\k*>\l*\relax\else
{\rotate(#2)\mov(#1,0){\sm*}}\loop
\ifnum\k*<\m*\advance\k*5{\rotate(\k*)\mov(#1,0){\sl*}}\repeat
{\rotate(#3)\mov(#1,0){\sl*}}\fi}}

\def\dasharc(#1)[#2,#3]{{\k**=#2\n*=#3\advance\n*-1\advance\n*-\k**
\L*=1000sp\L*#1\L* \multiply\L*\n* \multiply\L*\Nhalfperiods
\divide\L*57\N*\L* \divide\N*2000\ifnum\N*=0\N*1\fi
\r*\n*  \divide\r*\N* \ifnum\r*<2\r*2\fi
\m**\r* \divide\m**2 \l**\r* \advance\l**-\m** \N*\n* \divide\N*\r*
\k**\r* \multiply\k**\N* \dn*\n* 
\advance\dn*-\k** \divide\dn*2\advance\dn*\*one
\r*\l** \divide\r*2\advance\dn*\r* \advance\N*-2\k**#2\relax
\ifnum\l**<6{\rotate(#2)\mov(#1,0){\sm*}}\advance\k**\dn*
{\rotate(\k**)\mov(#1,0){\sl*}}\advance\k**\m**
{\rotate(\k**)\mov(#1,0){\sm*}}\loop
\advance\k**\l**{\rotate(\k**)\mov(#1,0){\sl*}}\advance\k**\m**
{\rotate(\k**)\mov(#1,0){\sm*}}\advance\N*-1\ifnum\N*>0\repeat
{\rotate(#3)\mov(#1,0){\sl*}}\else\advance\k**\dn*
\arc(#1)[#2,\k**]\loop\advance\k**\m** \r*\k**
\advance\k**\l** {\arc(#1)[\r*,\k**]}\relax
\advance\N*-1\ifnum\N*>0\repeat
\advance\k**\m**\arc(#1)[\k**,#3]\fi}}

\def\triangarc#1(#2)[#3,#4]{{\k**=#3\n*=#4\advance\n*-\k**
\L*=1000sp\L*#2\L* \multiply\L*\n* \multiply\L*\Nhalfperiods
\divide\L*57\N*\L* \divide\N*1000\ifnum\N*=0\N*1\fi
\d**=#2\Lengthunit \d*\d** \divide\d*57\multiply\d*\n*
\r*\n*  \divide\r*\N* \ifnum\r*<2\r*2\fi
\m**\r* \divide\m**2 \l**\r* \advance\l**-\m** \N*\n* \divide\N*\r*
\dt*\d* \divide\dt*\N* \dt*.5\dt* \dt*#1\dt*
\divide\dt*1000\multiply\dt*\magnitude
\k**\r* \multiply\k**\N* \dn*\n* \advance\dn*-\k** \divide\dn*2\relax
\r*\l** \divide\r*2\advance\dn*\r* \advance\N*-1\k**#3\relax
{\rotate(#3)\mov(#2,0){\sm*}}\advance\k**\dn*
{\rotate(\k**)\mov(#2,0){\sl*}}\advance\k**-\m**\advance\l**\m**\loop\dt*-\dt*
\d*\d** \advance\d*\dt*
\advance\k**\l**{\rotate(\k**)\rmov*(\d*,0pt){\sl*}}%
\advance\N*-1\ifnum\N*>0\repeat\advance\k**\m**
{\rotate(\k**)\mov(#2,0){\sl*}}{\rotate(#4)\mov(#2,0){\sl*}}}}

\def\wavearc#1(#2)[#3,#4]{{\k**=#3\n*=#4\advance\n*-\k**
\L*=4000sp\L*#2\L* \multiply\L*\n* \multiply\L*\Nhalfperiods
\divide\L*57\N*\L* \divide\N*1000\ifnum\N*=0\N*1\fi
\d**=#2\Lengthunit \d*\d** \divide\d*57\multiply\d*\n*
\r*\n*  \divide\r*\N* \ifnum\r*=0\r*1\fi
\m**\r* \divide\m**2 \l**\r* \advance\l**-\m** \N*\n* \divide\N*\r*
\dt*\d* \divide\dt*\N* \dt*.7\dt* \dt*#1\dt*
\divide\dt*1000\multiply\dt*\magnitude
\k**\r* \multiply\k**\N* \dn*\n* \advance\dn*-\k** \divide\dn*2\relax
\divide\N*4\advance\N*-1\k**#3\relax
{\rotate(#3)\mov(#2,0){\sm*}}\advance\k**\dn*
{\rotate(\k**)\mov(#2,0){\sl*}}\advance\k**-\m**\advance\l**\m**\loop\dt*-\dt*
\d*\d** \advance\d*\dt* \dd*\d** \advance\dd*1.3\dt*
\advance\k**\r*{\rotate(\k**)\rmov*(\d*,0pt){\sl*}}\relax
\advance\k**\r*{\rotate(\k**)\rmov*(\dd*,0pt){\sl*}}\relax
\advance\k**\r*{\rotate(\k**)\rmov*(\d*,0pt){\sl*}}\relax
\advance\k**\r*
\advance\N*-1\ifnum\N*>0\repeat\advance\k**\m**
{\rotate(\k**)\mov(#2,0){\sl*}}{\rotate(#4)\mov(#2,0){\sl*}}}}

\def\gmov*#1(#2,#3)#4{\rlap{\L*=#1\Lengthunit
\xL*=#2\L* \yL*=#3\L*
\rx* \gcos*\xL* \tmp* \gsin*\yL* \advance\rx*-\tmp*
\ry* \gcos*\yL* \tmp* \gsin*\xL* \advance\ry*\tmp*
\rx*=\xscale\rx* \ry*=\yscale\ry*
\xL* \the\cos*\rx* \tmp* \the\sin*\ry* \advance\xL*-\tmp*
\yL* \the\cos*\ry* \tmp* \the\sin*\rx* \advance\yL*\tmp*
\kern\xL*\raise\yL*\hbox{#4}}}

\def\rgmov*(#1,#2)#3{\rlap{\xL*#1\yL*#2\relax
\rx* \gcos*\xL* \tmp* \gsin*\yL* \advance\rx*-\tmp*
\ry* \gcos*\yL* \tmp* \gsin*\xL* \advance\ry*\tmp*
\rx*=\xscale\rx* \ry*=\yscale\ry*
\xL* \the\cos*\rx* \tmp* \the\sin*\ry* \advance\xL*-\tmp*
\yL* \the\cos*\ry* \tmp* \the\sin*\rx* \advance\yL*\tmp*
\kern\xL*\raise\yL*\hbox{#3}}}

\def\Earc(#1)[#2,#3][#4,#5]{{\k*=#2\l*=#3\m*=\l*
\advance\m*-6\ifnum\k*>\l*\relax\else\def\xscale{#4}\def\yscale{#5}\relax
{\angle**0\rotate(#2)}\gmov*(#1,0){\sm*}\loop
\ifnum\k*<\m*\advance\k*5\relax
{\angle**0\rotate(\k*)}\gmov*(#1,0){\sl*}\repeat
{\angle**0\rotate(#3)}\gmov*(#1,0){\sl*}\relax
\def\xscale{1}\def\yscale{1}\fi}}

\def\dashEarc(#1)[#2,#3][#4,#5]{{\k**=#2\n*=#3\advance\n*-1\advance\n*-\k**
\L*=1000sp\L*#1\L* \multiply\L*\n* \multiply\L*\Nhalfperiods
\divide\L*57\N*\L* \divide\N*2000\ifnum\N*=0\N*1\fi
\r*\n*  \divide\r*\N* \ifnum\r*<2\r*2\fi
\m**\r* \divide\m**2 \l**\r* \advance\l**-\m** \N*\n* \divide\N*\r*
\k**\r*\multiply\k**\N* \dn*\n* \advance\dn*-\k** \divide\dn*2\advance\dn*\*one
\r*\l** \divide\r*2\advance\dn*\r* \advance\N*-2\k**#2\relax
\ifnum\l**<6\def\xscale{#4}\def\yscale{#5}\relax
{\angle**0\rotate(#2)}\gmov*(#1,0){\sm*}\advance\k**\dn*
{\angle**0\rotate(\k**)}\gmov*(#1,0){\sl*}\advance\k**\m**
{\angle**0\rotate(\k**)}\gmov*(#1,0){\sm*}\loop
\advance\k**\l**{\angle**0\rotate(\k**)}\gmov*(#1,0){\sl*}\advance\k**\m**
{\angle**0\rotate(\k**)}\gmov*(#1,0){\sm*}\advance\N*-1\ifnum\N*>0\repeat
{\angle**0\rotate(#3)}\gmov*(#1,0){\sl*}\def\xscale{1}\def\yscale{1}\else
\advance\k**\dn* \Earc(#1)[#2,\k**][#4,#5]\loop\advance\k**\m** \r*\k**
\advance\k**\l** {\Earc(#1)[\r*,\k**][#4,#5]}\relax
\advance\N*-1\ifnum\N*>0\repeat
\advance\k**\m**\Earc(#1)[\k**,#3][#4,#5]\fi}}

\def\triangEarc#1(#2)[#3,#4][#5,#6]{{\k**=#3\n*=#4\advance\n*-\k**
\L*=1000sp\L*#2\L* \multiply\L*\n* \multiply\L*\Nhalfperiods
\divide\L*57\N*\L* \divide\N*1000\ifnum\N*=0\N*1\fi
\d**=#2\Lengthunit \d*\d** \divide\d*57\multiply\d*\n*
\r*\n*  \divide\r*\N* \ifnum\r*<2\r*2\fi
\m**\r* \divide\m**2 \l**\r* \advance\l**-\m** \N*\n* \divide\N*\r*
\dt*\d* \divide\dt*\N* \dt*.5\dt* \dt*#1\dt*
\divide\dt*1000\multiply\dt*\magnitude
\k**\r* \multiply\k**\N* \dn*\n* \advance\dn*-\k** \divide\dn*2\relax
\r*\l** \divide\r*2\advance\dn*\r* \advance\N*-1\k**#3\relax
\def\xscale{#5}\def\yscale{#6}\relax
{\angle**0\rotate(#3)}\gmov*(#2,0){\sm*}\advance\k**\dn*
{\angle**0\rotate(\k**)}\gmov*(#2,0){\sl*}\advance\k**-\m**
\advance\l**\m**\loop\dt*-\dt* \d*\d** \advance\d*\dt*
\advance\k**\l**{\angle**0\rotate(\k**)}\rgmov*(\d*,0pt){\sl*}\relax
\advance\N*-1\ifnum\N*>0\repeat\advance\k**\m**
{\angle**0\rotate(\k**)}\gmov*(#2,0){\sl*}\relax
{\angle**0\rotate(#4)}\gmov*(#2,0){\sl*}\def\xscale{1}\def\yscale{1}}}

\def\waveEarc#1(#2)[#3,#4][#5,#6]{{\k**=#3\n*=#4\advance\n*-\k**
\L*=4000sp\L*#2\L* \multiply\L*\n* \multiply\L*\Nhalfperiods
\divide\L*57\N*\L* \divide\N*1000\ifnum\N*=0\N*1\fi
\d**=#2\Lengthunit \d*\d** \divide\d*57\multiply\d*\n*
\r*\n*  \divide\r*\N* \ifnum\r*=0\r*1\fi
\m**\r* \divide\m**2 \l**\r* \advance\l**-\m** \N*\n* \divide\N*\r*
\dt*\d* \divide\dt*\N* \dt*.7\dt* \dt*#1\dt*
\divide\dt*1000\multiply\dt*\magnitude
\k**\r* \multiply\k**\N* \dn*\n* \advance\dn*-\k** \divide\dn*2\relax
\divide\N*4\advance\N*-1\k**#3\def\xscale{#5}\def\yscale{#6}\relax
{\angle**0\rotate(#3)}\gmov*(#2,0){\sm*}\advance\k**\dn*
{\angle**0\rotate(\k**)}\gmov*(#2,0){\sl*}\advance\k**-\m**
\advance\l**\m**\loop\dt*-\dt*
\d*\d** \advance\d*\dt* \dd*\d** \advance\dd*1.3\dt*
\advance\k**\r*{\angle**0\rotate(\k**)}\rgmov*(\d*,0pt){\sl*}\relax
\advance\k**\r*{\angle**0\rotate(\k**)}\rgmov*(\dd*,0pt){\sl*}\relax
\advance\k**\r*{\angle**0\rotate(\k**)}\rgmov*(\d*,0pt){\sl*}\relax
\advance\k**\r*
\advance\N*-1\ifnum\N*>0\repeat\advance\k**\m**
{\angle**0\rotate(\k**)}\gmov*(#2,0){\sl*}\relax
{\angle**0\rotate(#4)}\gmov*(#2,0){\sl*}\def\xscale{1}\def\yscale{1}}}

\newcount\CatcodeOfAtSign
\CatcodeOfAtSign=\the\catcode`\@
\catcode`\@=11
\def\@arc#1[#2][#3]{\rlap{\Lengthunit=#1\Lengthunit
\sm*\l*arc(#2.1914,#3.0381)[#2][#3]\relax
\mov(#2.1914,#3.0381){\l*arc(#2.1622,#3.1084)[#2][#3]}\relax
\mov(#2.3536,#3.1465){\l*arc(#2.1084,#3.1622)[#2][#3]}\relax
\mov(#2.4619,#3.3086){\l*arc(#2.0381,#3.1914)[#2][#3]}}}

\def\dash@arc#1[#2][#3]{\rlap{\Lengthunit=#1\Lengthunit
\d*arc(#2.1914,#3.0381)[#2][#3]\relax
\mov(#2.1914,#3.0381){\d*arc(#2.1622,#3.1084)[#2][#3]}\relax
\mov(#2.3536,#3.1465){\d*arc(#2.1084,#3.1622)[#2][#3]}\relax
\mov(#2.4619,#3.3086){\d*arc(#2.0381,#3.1914)[#2][#3]}}}

\def\wave@arc#1[#2][#3]{\rlap{\Lengthunit=#1\Lengthunit
\w*lin(#2.1914,#3.0381)\relax
\mov(#2.1914,#3.0381){\w*lin(#2.1622,#3.1084)}\relax
\mov(#2.3536,#3.1465){\w*lin(#2.1084,#3.1622)}\relax
\mov(#2.4619,#3.3086){\w*lin(#2.0381,#3.1914)}}}

\def\bezier#1(#2,#3)(#4,#5)(#6,#7){\N*#1\l*\N* \advance\l*\*one
\d* #4\Lengthunit \advance\d* -#2\Lengthunit \multiply\d* \*two
\b* #6\Lengthunit \advance\b* -#2\Lengthunit
\advance\b*-\d* \divide\b*\N*
\d** #5\Lengthunit \advance\d** -#3\Lengthunit \multiply\d** \*two
\b** #7\Lengthunit \advance\b** -#3\Lengthunit
\advance\b** -\d** \divide\b**\N*
\mov(#2,#3){\sm*{\loop\ifnum\m*<\l*
\a*\m*\b* \advance\a*\d* \divide\a*\N* \multiply\a*\m*
\a**\m*\b** \advance\a**\d** \divide\a**\N* \multiply\a**\m*
\rmov*(\a*,\a**){\unhcopy\spl*}\advance\m*\*one\repeat}}}

\catcode`\*=12

\newcount\n@ast

\def\n@ast@#1{\n@ast0\relax\get@ast@#1\end}
\def\get@ast@#1{\ifx#1\end\let\next\relax\else
\ifx#1*\advance\n@ast1\fi\let\next\get@ast@\fi\next}

\newif\if@up \newif\if@dwn
\def\up@down@#1{\@upfalse\@dwnfalse
\if#1u\@uptrue\fi\if#1U\@uptrue\fi\if#1+\@uptrue\fi
\if#1d\@dwntrue\fi\if#1D\@dwntrue\fi\if#1-\@dwntrue\fi}

\def\halfcirc#1(#2)[#3]{{\Lengthunit=#2\Lengthunit\up@down@{#3}\relax
\if@up\mov(0,.5){\@arc[-][-]\@arc[+][-]}\fi
\if@dwn\mov(0,-.5){\@arc[-][+]\@arc[+][+]}\fi
\def\lft{\mov(0,.5){\@arc[-][-]}\mov(0,-.5){\@arc[-][+]}}\relax
\def\rght{\mov(0,.5){\@arc[+][-]}\mov(0,-.5){\@arc[+][+]}}\relax
\if#3l\lft\fi\if#3L\lft\fi\if#3r\rght\fi\if#3R\rght\fi
\n@ast@{#1}\relax
\ifnum\n@ast>0\if@up\shade[+]\fi\if@dwn\shade[-]\fi\fi
\ifnum\n@ast>1\if@up\dshade[+]\fi\if@dwn\dshade[-]\fi\fi}}

\def\halfdashcirc(#1)[#2]{{\Lengthunit=#1\Lengthunit\up@down@{#2}\relax
\if@up\mov(0,.5){\dash@arc[-][-]\dash@arc[+][-]}\fi
\if@dwn\mov(0,-.5){\dash@arc[-][+]\dash@arc[+][+]}\fi
\def\lft{\mov(0,.5){\dash@arc[-][-]}\mov(0,-.5){\dash@arc[-][+]}}\relax
\def\rght{\mov(0,.5){\dash@arc[+][-]}\mov(0,-.5){\dash@arc[+][+]}}\relax
\if#2l\lft\fi\if#2L\lft\fi\if#2r\rght\fi\if#2R\rght\fi}}

\def\halfwavecirc(#1)[#2]{{\Lengthunit=#1\Lengthunit\up@down@{#2}\relax
\if@up\mov(0,.5){\wave@arc[-][-]\wave@arc[+][-]}\fi
\if@dwn\mov(0,-.5){\wave@arc[-][+]\wave@arc[+][+]}\fi
\def\lft{\mov(0,.5){\wave@arc[-][-]}\mov(0,-.5){\wave@arc[-][+]}}\relax
\def\rght{\mov(0,.5){\wave@arc[+][-]}\mov(0,-.5){\wave@arc[+][+]}}\relax
\if#2l\lft\fi\if#2L\lft\fi\if#2r\rght\fi\if#2R\rght\fi}}

\catcode`\*=11

\def\Circle#1(#2){\halfcirc#1(#2)[u]\halfcirc#1(#2)[d]\n@ast@{#1}\relax
\ifnum\n@ast>0\L*=\xscale\Lengthunit
\ifnum\angle**=0\clap{\vrule width#2\L* height.1pt}\else
\L*=#2\L*\L*=.5\L*\special{em:linewidth .001pt}\relax
\rmov*(-\L*,0pt){\sm*}\rmov*(\L*,0pt){\sl*}\relax
\special{em:linewidth \the\linwid*}\fi\fi}

\catcode`\*=12

\def\wavecirc(#1){\halfwavecirc(#1)[u]\halfwavecirc(#1)[d]}
\def\dashcirc(#1){\halfdashcirc(#1)[u]\halfdashcirc(#1)[d]}

\def\xscale{1}

\def\yscale{1}

\def\Ellipse#1(#2)[#3,#4]{\def\xscale{#3}\def\yscale{#4}\relax
\Circle#1(#2)\def\xscale{1}\def\yscale{1}}

\def\dashEllipse(#1)[#2,#3]{\def\xscale{#2}\def\yscale{#3}\relax
\dashcirc(#1)\def\xscale{1}\def\yscale{1}}

\def\waveEllipse(#1)[#2,#3]{\def\xscale{#2}\def\yscale{#3}\relax
\wavecirc(#1)\def\xscale{1}\def\yscale{1}}

\def\halfEllipse#1(#2)[#3][#4,#5]{\def\xscale{#4}\def\yscale{#5}\relax
\halfcirc#1(#2)[#3]\def\xscale{1}\def\yscale{1}}

\def\halfdashEllipse(#1)[#2][#3,#4]{\def\xscale{#3}\def\yscale{#4}\relax
\halfdashcirc(#1)[#2]\def\xscale{1}\def\yscale{1}}

\def\halfwaveEllipse(#1)[#2][#3,#4]{\def\xscale{#3}\def\yscale{#4}\relax
\halfwavecirc(#1)[#2]\def\xscale{1}\def\yscale{1}}

\catcode`\@=\the\CatcodeOfAtSign

\title{On the supersymmetric pseudo-QED}

\author{Van S\'ergio Alves} 
\email{vansergi@ufpa.br}

\affiliation{Faculdade de F\'{\i}sica, Universidade Federal do Par\'a,
	66075-110 Bel\'em, PA,  Brazil}

\author{M. Gomes}
\email{mgomes@if.usp.br}

\affiliation{Instituto de F\'\i sica, Universidade de S\~ao Paulo\\
	Caixa Postal 66318, 05315-970, S\~ao Paulo, SP, Brazil}

\author{A. Yu. Petrov}
\email{petrov@fisica.ufpb.br}

\affiliation{Departamento de F\'{\i}sica, Universidade Federal da Para\'{\i}ba\\
 Caixa Postal 5008, 58051-970, Jo\~ao Pessoa, Para\'{\i}ba, Brazil}

\author{A. J. da Silva}
\email{ajsilva@if.usp.br}

\affiliation{Instituto de F\'\i sica, Universidade de S\~ao Paulo\\
Caixa Postal 66318, 05315-970, S\~ao Paulo, SP, Brazil}

\begin{abstract}
Within the superfield approach, we discuss the three-dimensional supersymmetric  (SUSY) pseudo-QED. We prove that it is all-loop renormalizable. We demonstrate that the SUSY pseudo-QED action can be generated as a quantum correction from the coupling of a spinor gauge superfield to a set of $N$ massless complex scalar superfields. Afterwards, we calculate the two-point function of the scalar superfields in the pseudo-QED which displays a divergence vanishing in a certain gauge.
\end{abstract}

\maketitle

\section{Introduction}

The interest in three-dimensional  $3D$ models increased strongly after the discovery of the graphene \cite{Geim}. 
Graphene  is a two-dimensional system with the thickness of a single carbon atom organized in a honeycomb lattice with the electron obeying a  linear dispersion relation like a massless Dirac particle moving with the Fermi velocity.

The description of electronic interaction in two-dimensional systems such as graphene must take into account that the matter field is confined to the plane while the electromagnetic field is not. The solution to this problem was given in \cite{Marino}, where an effective description of electrons moving in a plane, but interacting through an electromagnetic field embedded in a four-dimensional space-time was obtained. Formally this is done by changing the usual term $\sim F^2_{\mu\nu}$ by another one, proportional to $F_{\mu\nu}(-\Box)^{-1/2}F^{\mu\nu}$, resulting in a nonlocal Maxwell-like term. This theory was called pseudo-quantum electrodynamics (PQED) or reduced quantum electrodynamics \cite{Gorbar} and, despite the nonlocality of the Maxwell term, the causality \cite{Amaral} and also unitarity   \cite{Unitarity} are respected. PQED has been used in the description of the electron-electron interactions in graphene \cite{PRX} and transition metal dichalcogenides (TMDs) \cite{Bandgap}, among other situations. 

We must stress that, since graphene's  discovery, there has been a great effort in the construction of a bridge between high-energy and condensed matter physics. The Klein paradox \cite{Kats}, the Zitterbewegung \cite{Kats2} and Schwinger's effect \cite{Allor} are well-known examples that provide a beautiful connection between them.

 Initially, it was believed that graphene was an effectively non-interacting system. However, experimental measurements of the fractional quantum Hall effect \cite{Du}, which is a typical feature of strongly correlated systems, and the renormalization of the Fermi velocity \cite{Elias} demonstrate that electromagnetic interactions are indeed important. Excellent reviews on this topic are presented in \cite{Neto}.

{From a theoretical point of view, it is relevant to find different ways to generate an effective theory with fractional power of the d'Alembertian operator, such as PQED. In this respect, recently \cite{pseudoqed1} it was shown that  under the bosonization process of free massive Dirac fermions, up to 1-loop order,  the usual Dirac Lagrangian ${\cal L}_D=\bar\psi(i\gamma^{\mu}\partial_{\mu}-m)\psi$ yields $\frac{1}{2}F_{\mu\nu}(-\Box)^{-1/2}F^{\mu\nu}+i(\frac{m}{8})\epsilon^{\mu\nu\alpha}A_{\mu}\partial_{\nu}(-\Box)^{-1/2}A_{\alpha}$, where $F_{\mu\nu}=\partial_{\mu}A_{\nu}-\partial_{\nu}A_{\mu}$, for $m^2\ll p^2$ regime, in the sense that the correlation functions of the free fermionic current are equal to the Green functions of the vector field $A_{\mu}$. This new Chern-Simons term, modified by the presence of the d'Alembertian operator, has the same classical symmetries as the usual Chern-Simons term i.e., it is gauge invariant and breaks the parity symmetry. Furthermore, the pole of the gauge field propagator occurs at $p^2=m^2$, similar to that of the fermionic field.  Another possibility is to consider these models in a more general field theory context which encompass supersymmetry.

Supersymmetry is a very powerful tool for the investigations of properties and possible applications of quantum field theory. In fact,
by describing bosons and fermions in a unique framework, it may  clarify fundamental aspects at very high energies as well to low energy
applications to condensed matter systems. In this respect, we would like to mention  works on the AdS/CFT correspondence \cite{maldacena}, noncommutative field theories \cite{noncom,cpn}, Lifshitz-like models, see f.e. \cite{Horava}.

Supersymmetry has been applied directly to the study of some systems in 2+1 
	dimensions \cite{cpn,FerTeix,3Dall,sugraph} (see also \cite{Petrov} and the references therein). In particular, graphene like systems
	have been investigated from this point of view. In the context of quantum mechanics, 
	supersymmetry has been used, for instance, to describe the quantum Hall effect in monolayer, bilayer and trilayer 
	graphene \cite{Ezawa}. This possible manifestation of SUSY in graphene (supersymmetric
	quantum mechanics) motivates the construction of a supersymmetric graphene field model.

 Clearly, a natural continuation of the above mentioned studies can consist in development of supersymmetric extension for models with negative fractional power
of the d'Alembertian operator. In this work we will show  that a supersymmetric version of pseudo-QED may arise  in the effective dynamics of a gauge superfield coupled to a scalar superfield. This  actually is very similar to what happens in the $CP^{N-1}$ model: classically the gauge field has no  dynamics which nevertheless emerges within the context of the $1/N$ expansion \cite{Dadda}.

The presence of the squared root of the d'Alembertian operator in the denominator of the Maxwell term
 ameliorates the infrared behavior of the model but worses the ultraviolet one. In spite of that, we found that the model is still ultraviolet renormalizable and also free of infrared divergences.  Besides this, we note that the "worsening" of the  ultraviolet asymptotics of the propagator, implies that the behaviour of quantum corrections in the theory becomes  more similar to that in the four-dimensional QED. From another side, for the pseudo-QED whose quadratic Lagrangian is ${\cal L}\propto F_{\mu\nu}(-\Box)^{-1/2}F^{\mu\nu}$, the Coulomb potential in the static limit is $V(r)\propto 1/r$, like in the usual four-dimensional case, see f.e. \cite{Unitarity}. In other words, the pseudo-QED allows to replay some important features of the four-dimensional situation in three-dimensional models.

The structure of the paper looks like follows. In the section II we define our model, demonstrate the arising of the pseudo-QED term as a quantum correction, and calculate the lower quantum contributions in the scalar sector. In the section III we present comments about the infrared behaviour of the SUSY pseudo-QED action. Our conclusions are presented in the section IV.

\section {Perturbative calculations}

The $3D$ pseudo-QED has been defined in \cite{pseudoqed} and its Action looks like ($\mu=0,1,2$):
\begin{eqnarray}
\label{pqed} 
S=\int d^3x \, [-\frac{1}{4}F^{\mu\nu}(-\Box)^{-1/2}F_{\mu\nu}+j^{\mu}_{\psi}A_{\mu}+{\cal L}_{\psi} \,], 
\end {eqnarray}
where ${\cal L_{\psi}}$ is the pure fermion matter Lagrangian, $A^{\mu}$ is the gauge potential, $F_{\mu \nu}=\partial_{\mu} A_{\nu}-\partial_{\nu} A_{\mu}$ is the gauge field strength and $j^{\mu}_{\psi}$ is the matter current. Its key feature is the presence of the $(-\Box)^{-1/2}$ making the action  nonlocal, and the ultraviolet  convergence to be worse than in the usual $3D$ QED, since the propagator behaves as $1/k$, where $k$ is the momentum. Let us discuss the superfield analogue of this theory, a scheme for its arising and perturbative behavior. 

As it is known, the supersymmetric $3D$ QED is described by the action ($\alpha=1,2$):
\begin{eqnarray}
\label{action0}
S&=&\int d^5 z [-\frac{1}{2} W^\alpha W_\alpha +j^{\alpha} A_{\alpha}+{\cal L}\, ],  
\end{eqnarray}
where
$W_\beta =\frac{1}{2}D^\alpha D_\beta A_\alpha$ 
is the spinor superfield strength constructed from the gauge spinor superpotential $A_\alpha$ (we follow the notation in \cite{Petrov} and the Appendix below), $j^{\alpha}$ is the spinor matter supercurrent and ${\cal L}$ is the pure supermatter Lagrangian.  As it is known, the component content of this theory contains the Lagrangian $-\frac{1}{4}F_{\mu\nu}F^{\mu\nu}+j^{\mu}_{\psi}A_{\mu}+{\cal L}_{\psi}$, besides terms dependent of the superpartners matter and gauge fields. Therefore, one can naturally conclude that to get the Lagrangian (\ref{pqed}) in the bosonic sector, one must start with the action
\begin{eqnarray}
\label{action}
S&=&\int d^5 z\,[-\frac{1}{2}W^\alpha (-\Box)^{-1/2}W_\alpha +j^{\alpha} A_{\alpha}+{\cal L}\,]. 
\end{eqnarray}
This is the $3D$ SUSY pseudo-QED action, being the natural superfield analogue of the pseudo-QED proposed in \cite{pseudoqed}. We note that, in principle, the presence of a negative degree of the d'Alembertian operator can be treated as a certain reminiscence of some nonlocal gravity models involving terms like $R\Box^{-1}R$, $R\Box^{-2}R$ \cite{Zhang:2016ykx,Fernandes:2017vvo}, which were studied mostly within the tree-level context.
 
 Our aims within this paper are, first, to explain a possible scheme for the emergence of this action, second, to study its possible perturbative impacts.

The first task is simpler. We start with the action of $N$ scalar massless superfields,  with the number $N$  assumed to be large to use the $1/N$ expansion (originally proposed within the quantum field theory context in \cite{Wilson}, see \cite{tHooft} for more detailed discussions and} \cite{Coleman:1980nk} for a review,
 \begin{eqnarray}
\label{acmat}
S_m&=&\int d^5 z
\Big[-\bar{\phi}_aD^2\phi_a+i\frac{g}{2} (\bar{\phi}_aA^\alpha D_\alpha \phi_a-
D_\alpha \bar{\phi}_aA^\alpha \phi_a)+\nonumber\\&+&
\frac{g^2}{2} \bar{\phi}_aA^\alpha A_\alpha\phi_a\Big]= \frac12\int d^{5}z (\overline{\nabla^{\alpha} \phi_{a}})\nabla_{\alpha} \phi_{a}\,,
\end{eqnarray}
where a sum over the repeated isotopic indices $a=1,\cdots,N$ is understood and the gauge supercovariant derivative $\nabla_{\alpha} = D_{\alpha}+i g A_{\alpha}$ was introduced. 
The free propagator of the scalar fields is
\begin{eqnarray}
<\bar{\phi}_a(-k,\theta_1)\phi_b(k,\theta_2)>=i\delta_{ab}\frac{D^2}{k^2}
\delta_{12}\,.
\end{eqnarray}
Actually this action is rather similar to the action of the SUSY $CP^{N-1}$ model (see the many references in \cite{cpn}, where a non commutative version of the SUSY $CP^{N-1}$ is studied). However, unlike \cite{cpn}, our theory is commutative, our scalar fields are massless (in order to get a pure $\sqrt{-\Box}$ factor with no mass dependence) and our model lacks the term involving a constraint in the modulus of $\Phi_a$, that is present in the definition of the $CP^{N-1}$ model.

So, we have the following contributions to the quadratic action of the gauge field (Fig. 1):

\vspace*{2mm}

\begin{figure}[htbp] 
	\begin{center} 
		\includegraphics[width={10cm}]{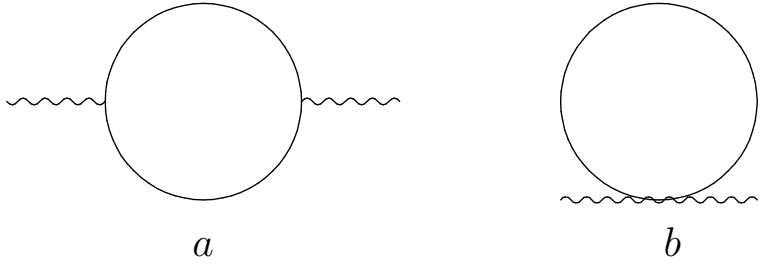}  
	\end{center} 
	\caption{Contributions to the two-point function of the gauge superfield.}
		\label{Fig:diagrams} 
\end{figure}

\vspace*{2mm}

Their contribution can be found in the same way as in \cite{cpn}, so here we merely quote the result,
\begin{eqnarray}
\label{stot}
\Sigma &=& \frac{Ng^2}{2} \int \frac{d^3p}{(2\pi)^3} d^2\theta \int \frac{d^3k}{(2\pi)^3}
\frac{k_{\gamma\beta}}{k^2(k+p)^2}
\nonumber\\&\times&
\Big[(D^2A^{\gamma}(-p,\theta)) A^{\beta}(p,\theta)
+\frac{1}{2} D^{\gamma}D^{\alpha}A_{\alpha}(-p,\theta) A^{\beta}(p,\theta)
\Big]\,.
\end{eqnarray}
Notice that, since our theory is commutative, this result may be obtained from  \cite{cpn} through the replacement of the phase factor $4\sin^2(k\wedge p)$ by $1$.
The above integral is evidently finite. It results in
\begin{eqnarray}
\Sigma=\frac{Ng^2}{4}\int d^2\theta\frac{d^3p}{(2\pi)^3} W^{\alpha}\frac{1}{\sqrt{p^2}}W_{\alpha},\label{action1}
\end{eqnarray}
which is nothing more than the first term of the desired action (\ref{action}).

Now, let us consider renormalizability of the theory whose action is formed by a sum of (\ref{acmat}) and (\ref{action1}) 

First of all, this theory is now given by the action
\begin{eqnarray}
	\label{theory}
S&=&\int d^5z \frac{Ng^2}{4}W^{\alpha}\frac{1}{\sqrt{-\Box}}W_{\alpha}+\nonumber\\
&+&\int d^5 z
\Big[-\bar{\phi}_aD^2\phi_a+i\frac{g}{2} (\bar{\phi}_aA^\alpha D_\alpha \phi_a-
D_\alpha \bar{\phi}_aA^\alpha \phi_a)+\nonumber\\&+&
\frac{g^2}{2} \bar{\phi}_aA^\alpha A_\alpha\phi_a\Big],
\end{eqnarray}
with the additional proviso that diagrams that contain the graphs shown before as subgraphs must be discarded, as it is required by the $1/N$ expansion formalism.
We can fix the gauge by adding the term
\begin{eqnarray}
	\label{gf}
S_{gf}=N\frac{1}{\xi}\int d^5z 
(D^{\alpha}A_{\alpha})(\frac{g^2}{16}\frac{D^2}{\sqrt{-\Box}})(D^{\beta}A_{\beta}).
\end{eqnarray}
The gauge propagator will asymptotically behave in ultraviolet limit as $\frac{1}{k}$, as well as the matter one.

So, the superficial degree of divergence (SDD) is
\begin{equation}
\omega=2L+\frac{1}{2}V_3-P,
\end{equation} 
where $V_3$ is a number of triple vertices and $P$ is the number of propagators.  Each loop contributes $2$ as it is usual in three-dimensional superfield theories (this follows from the fact that for each loop two D's must be used to shrink a loop into a point through the identity $\delta_{12}D^2\delta_{12}=\delta_{12}$). Taking into account structure of vertices we find
\begin{equation}
\omega=2-\frac{E}{2}-\frac{N_D}{2},
\end{equation}
where $E$ is a number of external legs, and we took into account as usual the number of spinor supercovariant derivatives transferred to external legs, denoted here as $N_D$. We see that the theory is renormalizable. Notice that the corrections with three or four gauge legs and no scalar legs are forbidden by the gauge symmetry, so, the only divergences are corrections to kinetic terms and just the scalar-vector triple and quartic vertices that can be read off from the classical action (\ref{acmat}). Actually, the gauge symmetry requires the divergences to be proportional to $(\overline{\nabla^{\alpha}\phi})\nabla_{\alpha}\phi$ and $\phi^*_a\phi_a$ in the scalar sector, and $A^{\alpha}W_{\alpha}$ in the gauge sector (nonlocal terms involving  more derivatives of $A_{\alpha}$ and hence, by dimensional reasons, negative degrees of $\Box$ should be finite since any negative degree of $\Box$ can be generated only by a finite integral over moments). 
	
	Therefore, our next step consists in finding the contributions in the scalar sector (Fig. 2). We will obtain the $D^{\alpha}\phi_aD_{\alpha}\bar{\phi}_a$ contribution which, using the gauge symmetry, can be naturally completed  to $(\overline{\nabla^{\alpha}\phi})\nabla_{\alpha}\phi$ contribution.
It is given by the supergraphs $c, d$.

\vspace*{2mm}

\begin{figure}[htbp] 
	\begin{center} 
		\includegraphics[width={10cm}]{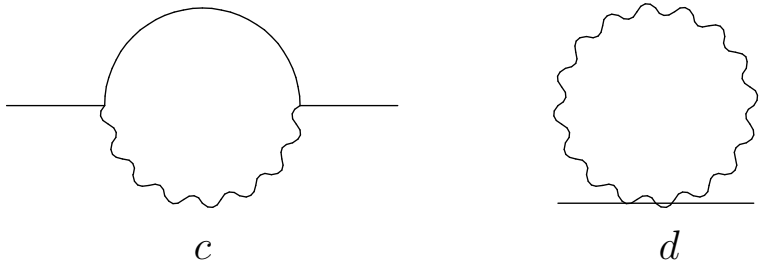}  
	\end{center} 
	\caption{Contributions to the two-point function of the matter superfield.}
\end{figure}

\vspace*{2mm}

To proceed with this calculation, we can apply the results obtained for the noncommutative  $CP^{N-1}$ model in the fundamental representation \cite{FerTeix} (where, unlike the adjoint representation \cite{cpn}, there is no phase factors), with  the only differences  that now we have no mass term, and the Lagrange multiplier field is absent. It is easy to see that if the quadratic action, with the gauge fixing additive term, being the sum of the gauge sectors of (\ref{theory}) and (\ref{gf}), looks like
\begin{eqnarray}
	S=-\frac{N}{4}\int d^5z A_{\alpha}f(\Box)D^2(D^{\beta}D^{\alpha}+\frac{1}{\xi}D^{\alpha}D^{\beta})A_{\beta},
	\end{eqnarray}
and the propagator is
\begin{eqnarray}
	<A_{\beta}(z_1)A_{\gamma}(z_2)>=\frac{i}{2}\frac{D^2}{\Box^2 N f(\Box)}(D_{\gamma}D_{\beta}+\xi D_{\beta}D_{\gamma})\delta^5(z_1-z_2).
\end{eqnarray}
In our case, $f(\Box)=\frac{g^2}{4\sqrt{-\Box}}$. We note at the same time that, unlike the $CP^{N-1}$ model, there is no auxiliary scalar superfield in our theory since the pseudo-QED does not involve any constraints. 


It is clear that the (d) diagram is $\Gamma_d=\frac{g^2}{2}\int d^5z \bar{\phi}_a\phi_a<A^{\alpha}A_{\alpha}>$. Then, we use the well-known formula $D_{\alpha}D_{\beta}=i\partial_{\alpha\beta}-C_{\alpha\beta}D^2$. Taking into account that we have the additional $D^2$ factor in the propagator, we find that our contribution is proportional to $(i\partial_{\alpha\beta}D^2-C_{\alpha\beta}\Box)\delta^5(z_1-z_2)|_{z_1=z_2}$, which vanishes explicitly (the second term yields a zero trace in the Grassmannian sector, and the first one, after the Fourier transform, yields the integral propotional to $\int d^3k k_{\mu}h(k^2)$, which vanishes by symmetry reasons). We note that the analogous contribution vanishes in the zero mass limit also in the $CP^{N-1}$ model \cite{cpn,FerTeix}.

So, we rest with the contribution in  Fig. (c). 
It looks like
\begin{eqnarray}
\Gamma_c&=&\frac{g^2}{2}\int d^{5}1\,d^{5}2\,<A^{\alpha}(1)A^{\beta}(2)>\times\nonumber\\&\times&
\Big[
D_{\alpha}\phi(1)\bar{\phi}(2)<\bar{\phi}(1)D_{\beta}\phi(2)>+<D_{\alpha}\phi(1)\bar{\phi}(2)>\bar{\phi}(1)D_{\beta}\phi(2)-\nonumber\\&-&<D_{\alpha}\bar{\phi}(1)D_{\beta}\phi(2)>\phi(1)\bar{\phi}(2)-D_{\alpha}\bar{\phi}(1)D_{\beta}\phi(2)<\phi(1)\bar{\phi}(2)>-\nonumber\\
&-&\bar{\phi}(1)\phi(2)<D_{\alpha}\phi(1)D_{\beta}\bar{\phi}(2)>-<\bar{\phi}(1)\phi(2)>D_{\alpha}\phi(1)D_{\beta}\bar{\phi}(2)+\nonumber\\&+&
<D_{\alpha}\bar{\phi}(1)\phi(2)>\phi(1)D_{\beta}\bar{\phi}(2)+D_{\alpha}\bar{\phi}(1)\phi(2)<\phi(1)D_{\beta}\bar{\phi}(2)>
\Big].
\end{eqnarray}
To do the calculation, we can employ the results obtained in \cite{FerTeix}, restricting ourselves to the zero mass case. Actually, the only difference between our  effective propagator and that one from \cite{FerTeix} treated at $m=0$, is the slight modification of the constant multiplier in the $f(\Box)$ function (in our case, it looks like $\frac{g^2}{8}$ while in \cite{FerTeix}, the factor is $\frac{1}{8}$ since in our case the couplings are introduced in gauge-matter vertices instead of the $\frac{N}{g}$ term absent here). Adopting the results of \cite{FerTeix}, we find
\begin{eqnarray}
	\label{ir}
\Gamma_c=-\frac{g^2(\xi+2)}{2N}\int d^2\theta\int\frac{d^3p}{(2\pi)^3}D^{\alpha}\phi(-p)D_{\alpha}\overline{\phi}(p)\int\frac{d^3k}{(2\pi)^3}\frac{1}{\sqrt{k^2}(k+p)^2},
\end{eqnarray}
which diverges logarithmically but vanishes at $\xi=-2$, just as the divergent part of the contribution to the two-point function of the scalar field from the same graph in \cite{FerTeix} does.

 In principle, the gauge symmetry allows to expect that the lower contributions to three- and four-point functions within the $1/N$ expansion, with use of the dimensional regularization scheme, with the space-time dimension is promoted to $d=3-\epsilon$,  will yield a result like
	$\Gamma_{\phi,A}\propto\frac{1}{N\epsilon}(\overline{\nabla^{\alpha}\phi})\nabla_{\alpha}\phi$.

\section{A comment on the behavior of the theory in the infrared limit}

Since our pseudo-QED is an essentially massless theory, it is interesting to study the problem of infrared singularities whose presence is a relevant characteristic for massless theories. Here, it is impossible to follow the methodology developed in \cite{irQED} for the non-supersymmetric theory since in the superfield case there is no explicit Dirac matrices in the propagators of any field. However, there is one interesting observation  that we can do on the basis of our result (\ref{ir}). Indeed, the $\Gamma_c$ (\ref{ir}), after the dimensional regularization, integration and the  subtraction of the divergence with an appropriate counterterm, behaves as $\Gamma_c\propto \frac{g^2(\xi+2)}{N}\int d^2\theta\int \frac{d^3p}{(2\pi)^3} \phi D^2\bar{\phi}\ln\frac{p^2}{\mu^2}$. We note that this expression is written in terms of the superfields, so, we project it to the component fields.  We recall that,   in the notation of \cite{Petrov}, the  component content of $\phi$ is
\begin{eqnarray}
\phi=\varphi+\theta^{\alpha} \psi_{\alpha}-\frac12 \theta^{\alpha}\theta_{\alpha}F.
\end{eqnarray}
For the scalar sector, we obtain $\Gamma_c\propto \varphi (p) p^2\ln\frac{p^2}{\mu^2}\varphi(-p)$ which does not display any infrared singularities due to $p^2$ present in the kinetic term. A similar situation occurs also in the Dirac sector where we have $\bar{\psi}\gamma^{\mu}p_{\mu}\psi\ln\frac{p^2}{\mu^2}$. The only infrared problem can arise in the auxiliary field sector \cite{Petrov}, where we have a result proportional to $F(-p)\ln\frac{p^2}{\mu^2}F(p)$, however, as follows from (\ref{ir}), even this term vanishes in a certain gauge. We conclude that the infrared problem can be ruled out under an appropriate gauge condition. 
 It is interesting to note that, if we had a self-coupling $\lambda\Phi^3$, one could  solve the effective equations of motion for the $F$ field, so that the solution for $F$, up to some numerical coefficients, will yield $F=-(1+\ln\frac{\Box}{\mu^2})^{-1}(\lambda\phi^2)$, i.e. in the infrared limit the denominator becomes large, hence the solution for $F$ does not blow up, being suppressed instead. Thus, even in this case the additive term $F\ln\frac{p^2}{\mu^2}F$ does not generate new dangerous ghost-like degrees of freedom.

\section{Conclusions}

We considered the three-dimensional SUSY pseudo-QED.  We demonstrated explicitly 
 that our theory is renormalizable. In this theory, renormalization properties are worse than in the usual 3D super-QED known to be super-renormalizable {due to its superficial degree of divergence, but even finite in a certain gauge \cite{3Dall}. We note that the theory is essentially unitary since its bosonic counterpart is unitary \cite{Unitarity}. In principle, this is natural because we have no higher derivatives in the action. We note that one can expect that the complete leading divergent contribution in the $1/N$ expansion looks like $\Gamma_{\phi,A}\propto\frac{1}{N\epsilon}(\overline{\nabla^{\alpha}\phi_a})\nabla_{\alpha}\phi_a$.  To close the paper, we conclude that these results can be applied to study of supersymmetric graphene models being the further development of the theory proposed in \cite{sugraph}. Also, it is natural to expect arising of some analogues of these results within studies of nonlocal gravity involving negative degrees of the d'Alembertian operator.

{\bf Acknowledgments.}
 The work of A. J. S.  was partially supported by Conselho Nacional de
Desenvolvimento Cient\'\i fico e Tecnol\'ogico (CNPq), project No. 309915/2021-0 . A. Yu. P. has been partially supported by the CNPq project No. 301562-2019/9. 
	V. S. Alves bas been partially supported by CAPES/NUFFIC, finance code 0112.
	\appendix
\section{The  $2+1$ dimensional superspace}

 The three-dimensional superspace consists of the usual three-dimensional coordinates and  one spinor Grassmanian variable, $\theta_{\alpha}$, $\alpha=1,2$. Spinorial indices are raised and lowered by the Hermitian symbol $C_{\alpha\beta}=-i \epsilon_{\alpha\beta}=\left(\begin{array}{cc}0&-i\\i&0
 \end{array}\right)$.
  By introducing the Dirac matrices
$(\gamma^{0})^{\alpha}_{\phantom a \beta}=-i\sigma_{2}$, $(\gamma^{1})^{\alpha}_{\phantom a \beta}=\sigma_{1}$, $(\gamma^{2})^{\alpha}_{\phantom a \beta}=\sigma_{3}$, where $\sigma_{i}$, $i=1,2,3$ are the Pauli matrices, we may write supersymetry transformations as
generated by the operator
\begin{eqnarray}
Q = i\partial_{\alpha} + \theta^{\beta}\partial_{\beta\alpha},
\end{eqnarray}
where $\partial_{\beta\alpha}\equiv \gamma^{\mu}_{\beta\alpha}\partial_{\mu}$.  These generators satisfy the relation $\{Q_{\alpha},Q_{\beta}\}=2i\partial_{\alpha\beta}$. Here and further, for any spinor $B^{\alpha}$, we denote $B^2\equiv\frac{1}{2}B^{\alpha}B_{\alpha}$.

 Superfields are functions in the superspace which under a supersymmetry operation transforms as
\begin{eqnarray}
\Phi \qquad\rightarrow\qquad e^{i\eta^\alpha Q_{\alpha}}\Phi
\end{eqnarray}
The simplest example of a superfield is a scalar superfield which
may be expanded  as
\begin{eqnarray}
\phi(x,\theta)=\varphi(x)+ \theta^{\alpha}\psi_{\alpha}-\theta^{2}F(x),
\end{eqnarray}
where, under a Lorentz transformation, $\varphi$ and $F$ are scalar fields while $\psi$ is a  Lorentz spinor.
Linear combinations of superfields are superfields but their spinorial derivatives are not. In this situation,  we may use a supercovariant derivative, defined as
\begin{eqnarray}
D_{\alpha} = \partial_{\alpha}+i \theta^{\beta}\partial_{\beta\alpha},
\end{eqnarray}
defined to commute with the SUSY generators, $\{Q_{\alpha},D_{\beta}\}=0$. The supercovariant derivatives display several important properties:
\begin{eqnarray}
\{D_{\alpha},D_{\beta}\}=2i\partial_{\alpha\beta}; \quad\, [D_{\alpha},D_{\beta}]=-2C_{\alpha\beta}D^2;\quad\, D^{\alpha}D_{\beta}D_{\alpha}=0;\quad\, \{D_{\alpha},D^2\}=0.
\end{eqnarray}
Another important superfield is the spinor one, whose component form is
\begin{eqnarray}
A_{\alpha}(x,\theta)&=&\chi_{\alpha}(x)-\theta_{\alpha}B(x)+i\theta^{\beta}V_{\beta\alpha}(x)
-2\theta^2[\lambda_{\alpha}(x)+\frac{i}{2}\partial_{\alpha\beta}\chi^{\beta}(x)],
\end{eqnarray}
The importance of this field consists in the fact that it allows to introduce the gauge invariance within the superfield framework. Indeed, the object 
\begin{equation}
W_{\alpha}=\frac{1}{2}D^{\beta}D_{\alpha}A_{\beta},
\end{equation}
called the superfield strength, is invariant under gauge transformations $\delta A_{\beta}=D_{\beta}\xi$. This strength is used to construct actions of superfield QED and Chern-Simons theories.


\begin{thebibliography}{10}
\bibitem{Geim} K. S. Novoselov et al., Science 306 (2004) 666; 
Nature 438 (2005) 197; A. K. Geim and K. S. Novoselov, 
Nat. Mater. 6 (2007) 183; 
M. I.  Katsnelson, Materials Today 10 (2007) 20.

\bibitem{Marino}  E. C. Marino, Nucl. Phys. B408, 551 (1993).

\bibitem{Gorbar} E. V. Gorbar, V. P. Gusynin, and V. A. Miransky, Phys. Rev. D64, 105028 (2001).

\bibitem{Amaral}  R. L. P. G. do Amaral and E. C. Marino, 
J. Phys. A25, 5183 (1992).

\bibitem{Unitarity}  E. C. Marino, L. O. Nascimento, V. S. Alves, and C. Morais
Smith, 
Phys. Rev. D90, 105003 (2014).


\bibitem{PRX} E. C. Marino, L. O. Nascimento, V. S. Alves, and C. M. Smith, Phys. Rev. X 5, 011040 (2015); N. Menezes, V. S. Alves, E. C. Marino, L. O. Nascimento, and C. Morais Smith, Phys. Rev. B 95, 245138 (2017).

\bibitem{Bandgap} E. C. Marino, L. O. Nascimento, V. S. Alves, N. Menezes, and C. Morais Smith, 2D Mater. 5, 041006 (2018); L. Fernandez, V. S. Alves, L. O. Nascimento, F. Pena, M. Gomes, and E. C. Marino, Phys. Rev. D 102, 016020 (2020).

\bibitem{Kats} M. I. Katsnelson, K. S. Novoselov, and A. K. Geim, 
Nat. Phys. 2, 620 (2006).

\bibitem{Kats2} M. I. Katsnelson, 
Eur. Phys. J. B 51, 157 (2006); J. Tworzydlo, B. Trauzettel, M. Titov, A. Rycerz, and C. W. J. Beenakker, 
Phys. Rev. Lett. 96, 246802 (2006).

\bibitem{Allor} D. Allor, T. D. Cohen, and D. A. McGady, 
Phys. Rev. D 78, 096009 (2008).

\bibitem{Du} X. Du, I. Skachko, F. Duerr, A. Luican, and E. Y. Andrei, 
Nature (London) 462, 192 (2009);  K. l. Bolotin, F. Ghahari, M. D. Shulman, H. L. Stormer, and P. Kim, 
Nature (London) 462, 196 (2009); F. Ghahari, Y. Zhao, P. Cadden-Zimansky, K. Bolotin, and P. Kim, 
Phys. Rev. Lett. 106, 046801 (2011).

\bibitem{Elias} D. C. Elias, R. V. Gorbachev, A. S. Mayorov, S. V. Morozov, A. A. Zhukov, P. Blake, L. A. Ponomarenko, I. V. Grigorieva, K. S. Novoselov, F. Guinea,  A. K. Geim, 
Nat. Phys. 7, 701 (2011); J. Chae, S. Jung, A. F. Young, C. R. Dean, L. Wang, Y. Gao, K. Watanabe, T. Taniguchi, J. Hone, K. L. Shepard, P. Kim, N. B. Zhitenev, J. A. Stroscio, 
Phys. Rev. Lett. 109, 116802 (2012).

\bibitem{Neto} A. H. Castro Neto, F. Guinea and N. M. R. Peres, 
Phys. World 19 (2006) 33; M. I. Katsnelson, 
Mater. Today 10 (2007) 20; A. K. Geim and A. H. MacDonald, 
Phys. Today 60 (2007) 35; A. K. Geim and K.S. Novoselov, 
Nature Mater. 6 (2007) 183; A. H. Castro Neto, F. Guinea, N. M. R. Peres, K. S. Novoselov and A. K. Geim, 
Rev. Mod. Phys. 81 (2009) 109; A. K. Geim, 
Science 324 (2009) 1530.

\bibitem{pseudoqed1} G.~C.~Magalh\~aes, V.~S.~Alves,  L.~O.~Nascimento and, E.~C.~Marino, Phys.\ Rev.\ D103, 116022 (2021)


\bibitem{maldacena}J. Maldacena, Adv. Theor. Math. Phys., 2, 231 (1998); S. Gubser,   I. Klebanov, A.  Polyakov, Phys. Lett. B. 428, 105 (1998); E. Witten, Nucl. Phys. B. 311, 46 (1998).

\bibitem{noncom} H. O. Girotti, M. Gomes, V. O. Rivelles, A. J. Da Silva,
Nucl. Phys. B 587, 299 (2000).

  
\bibitem{cpn} E.~A.~Asano, H.~O.~Girotti, M.~Gomes, A.~Y.~Petrov, A.~G.~Rodrigues and A.~J.~da Silva,
  Phys.\ Rev.\ D69, 105012 (2004).
  
  \bibitem{Horava} C.~F.~Farias, M.~Gomes, J.~R.~Nascimento, A.~Y.~Petrov and A.~J.~da Silva,
  Phys. Rev. D85, 127701 (2012);
  C.~F.~Farias, M.~Gomes, J.~R.~Nascimento, A.~Y.~Petrov and A.~J.~da Silva,
  Phys. Rev. D89, 025014 (2014);
  M. Gomes, J. R. Nascimento, A. Yu. Petrov, A. J. da Silva, Phys. Rev. D 90, 125022 (2014).
  

\bibitem{3Dall} A.~F.~Ferrari, M.~Gomes, A.~C.~Lehum, A.~Y.~Petrov and A.~J.~da Silva,
  Phys.\ Rev.\ D77, 065005 (2008).

\bibitem{sugraph} E.~M.~C.~Abreu, M.~A.~De Andrade, L.~P.~G.~De Assis, J.~A.~Helayel-Neto, A.~L.~M.~A.~Nogueira and R.~C.~Paschoal,
  JHEP 1105, 001 (2011).
  \bibitem{FerTeix}  A.~F.~Ferrari, A.~C.~Lehum, A.~J.~da Silva and F.~Teixeira,
J.\ Phys.\ A40, 7803 (2007).
  
  \bibitem{Petrov} A. Yu. Petrov, {\it Quantum Superfield Supersymmetry}, Fundam.Theor. Phys. 202, Springer, 2021,  hep-th/0106094.
  
 \bibitem{Ezawa} M. Ezawa, 
Physica E 40, 269 (2007); M. Ezawa, 
 Phys. Lett. A 372, 924 (2008).
  
  
\bibitem{Dadda} A. D'Adda,  M. L\"uscher,  P. Di  Vecchia,  Nucl. Phys. B146, 63 (1978).  
    
\bibitem{pseudoqed} G.~C.~Magalhães, V.~S.~Alves, E.~C.~Marino and L.~O.~Nascimento,
  Phys.\ Rev.\ D101, 116005 (2020).


 \bibitem{Zhang:2016ykx} Y.-li.~Zhang, K.~Koyama, M.~Sasaki and G.-B.~Zhao,
JHEP 03, 039 (2016).

\bibitem{Fernandes:2017vvo}
K.~Fernandes and A.~Mitra,
Phys. Rev. D97, 105003 (2018).

 \bibitem{Wilson} K. Wilson, Phys. Rev. D7, 2911 (1973).
 
 \bibitem{tHooft} G. 't Hooft, Nucl. Phys. B72, 461 (1974). 
  
  \bibitem{Coleman:1980nk} S. R. Coleman, "1/N".  In {\it Aspects of Symmetry: Selected Erice Lectures}, pp. 351-402, Cambridge University Press, 1985.
  


\bibitem{irQED} H. O. Girotti, M. Gomes, A. J. da Silva, Mod. Phys. Lett. A9, 2699 (1994).



\end{thebibliography}
\end{document}